\def\draftversion{false}
\RequirePackage{ifthen}
\ifthenelse{\equal{\draftversion}{true}}{\documentclass[aps,prx,10pt,galley,amsmath,amssymb,showpacs,
                 superscriptaddress]{revtex4}
}
{
  \documentclass[aps,prx,10pt,twocolumn,showpacs,
                 superscriptaddress]{revtex4}
}

\usepackage{times}
\usepackage{amsmath}
\usepackage{amssymb}
\usepackage{graphicx}
\usepackage[usenames, dvipsnames]{color} 
\usepackage{bm} 
\usepackage{subfigure}
\newcommand{\angstrom}{\mbox{\normalfont\AA}}

\def\I{\uppercase\expandafter{\romannumeral 1}}
\def\II{\uppercase\expandafter{\romannumeral 2}}
\def\III{{\uppercase\expandafter{\romannumeral 3}}}
\def\IV{{\uppercase\expandafter{\romannumeral 4}}}
\def\V{{\uppercase\expandafter{\romannumeral 5}}}
\def\VI{{\uppercase\expandafter{\romannumeral 6}}}
\def\VII{{\uppercase\expandafter{\romannumeral 7}}}
\def\i{\lowercase\expandafter{\romannumeral 1}}
\def\ii{\lowercase\expandafter{\romannumeral 2}}
\def\iii{{\lowercase\expandafter{\romannumeral 3}}}
\def\iv{{\lowercase\expandafter{\romannumeral 4}}}
\def\v{{\lowercase\expandafter{\romannumeral 5}}}
\def\vi{{\lowercase\expandafter{\romannumeral 6}}}
\def\vii{{\lowercase\expandafter{\romannumeral 7}}}

\def\nn{\nonumber\\}

\def\uB{\mu_{\textrm{B}}}
\def\mnsn{Mn$_3$Sn}
\def\rr{\mathbf{r}}

\def\Green#1{\textcolor{OliveGreen}{#1}}

\begin{document}

\title{Anomalous Hall effect and topological defects in antiferromagnetic Weyl semimetals: Mn$_3$Sn/Ge}

\author{Jianpeng Liu}
\affiliation{ Kavli Institute for Theoretical Physics, University of California, Santa Barbara
CA 93106, USA}

\author{Leon Balents}
\affiliation{ Kavli Institute for Theoretical Physics, University of California, Santa Barbara
CA 93106, USA}

\date{\today}

\begin{abstract}
We theoretically study the interplay between bulk Weyl electrons and magnetic 
topological defects, including magnetic domains, domain walls, and $\mathbb{Z}_6$ vortex lines, 
in the  antiferromagnetic Weyl semimetals \mnsn\ and Mn$_3$Ge with
negative vector chirality.   
We argue that these materials possess a hierarchy of energies scales which allows a
description of the spin structure and spin dynamics using a XY
  model with $\mathbb{Z}_6$ anisotropy.  We propose a dynamical
equation of motion for the XY order parameter, which implies 
the presence of $\mathbb{Z}_6$ vortex lines, the double-domain pattern in the
presence of magnetic fields, and the
ability to control domains with current.  We also introduce a
minimal electronic model which allows efficient calculation of the
electronic structure in the antiferromagnetic
configuration, unveiling Fermi arcs at domain walls, and sharp
quasi-bound states at $\mathbb{Z}_6$ vortices.  Moreover, we have shown how these
materials may allow electronic-based imaging of antiferromagnetic
microstructure, and propose a possible device based on
domain-dependent anomalous Hall effect.

\end{abstract}


\maketitle

\def\scr{\scriptsize}
\ifthenelse{\equal{\draftversion}{true}}{
  \marginparwidth 2.7in
  \marginparsep 0.5in
  \newcounter{comm} 
  \def\commnext{\stepcounter{comm}}
  \def\commtext{{\bf\color{blue}[\arabic{comm}]}}
  \def\commmar{{\bf\color{blue}[\arabic{comm}]}}
  \def\lbm#1{\commnext\marginpar{\small LB\commmar: #1}\commtext}
  \def\jlm#1{\commnext\marginpar{\small JPL\commmar: #1}\commtext}
  \def\mlab#1{\marginpar{\small\bf #1}}
  \def\tnewpage{\newpage\marginpar{\small Temporary newpage}}
  \def\tfootnote#1{\Green{\scr [FOOTNOTE: #1]}}
}{
  \def\lbm#1{}
  \def\jlm#1{}
  \def\mlab#1{}
  \def\tnewpage{}
  \def\tfootnote#1{\footnote{#1}}
}


The Hall effect has long been a nucleation center for geometry and
topology in the physics of solids.  In the 1950s, prescient work of
Karplus and Luttinger identified Berry curvature of electron
wavefunctions as the heart of the anomalous Hall effect (AHE) in
ferromagnets\cite{karplus-pr54,ahe-rmp10}.  In the 1980s, topology entered with the discovery of the
quantum Hall effect.  These ideas came together in the mid-2000s to
unveil broad applications to electronic systems in the form of
topological insulators, superconductors \cite{kane-rmp10,zhang-rmp11}
and semimetals with topological Weyl (and other) fermion excitations
\cite{wan-prb11,burkov-prb11,burkov-prl11,xu-prl11,murakami-njphys-07,na3bi-theory,cd3as2-theory,turner-13-review,jpliu-prb14-a,taas-theory-dai,typeii-wsm,taas-exp-chen,taas-exp-ding1,taas-exp-hasan}.
The AHE re-appears as one of the key emergent
properties of topological semimetals, and coming full circle, most 
ferromagnets are now believed to host Weyl fermions to which their AHE
is at least in part attributed.

The dissipationless nature of the Hall effect also makes it
interesting for applications.  Uses based on ferromagnets may,
however, be limited by the difficulty of miniaturization posed by large
fields generated by the magnetization.  For this reason,
antiferromagnetic realizations of AHE may be of practical interest,
but the microstructure, dynamics, and AHE of antiferromagnets are
relatively uninvestigated.  Here we attack these issues in the family of noncollinear
antiferromagnets including  Mn$_3$Sn and Mn$_3$Ge, for which a strong
AHE was predicted and then experimentally verified to exist\cite{chen-prl14,
  mn3sn-exp, mn3ge-exp}.  First principles calculations further
indicate that in Mn$_3$Sn and Mn$_3$Ge there are Weyl nodes around the
Fermi level \cite{yan-mn3sn-arxiv16, felser-epl14}.  We argue that
these materials possess a hierarchy of energies scales which permits a
description of the microstructure and spin dynamics as an {\em XY
  model with $\mathbb{Z}_6$ anisotropy}.  We propose a dynamical
equation of motion for the XY order parameter, which implies a rich
domain structure, the presence of $\mathbb{Z}_6$ vortex lines, and the
ability to control domains with current.  We further introduce a
minimal electronic model which allows efficient calculation of the
electronic structure in a {\em textured} antiferromagnetic
configuration, unveiling Fermi arcs at domain walls, and sharp
quasi-bound states at $\mathbb{Z}_6$ vortices.  We show how these
materials may allow electronic-based imaging of antiferromagnetic
microstructure (difficult to observe magnetically due to the
lattice-scale variations) and propose a possible device based on
domain-dependent AHE.  

\begin{figure}
\includegraphics[width=3.4in]{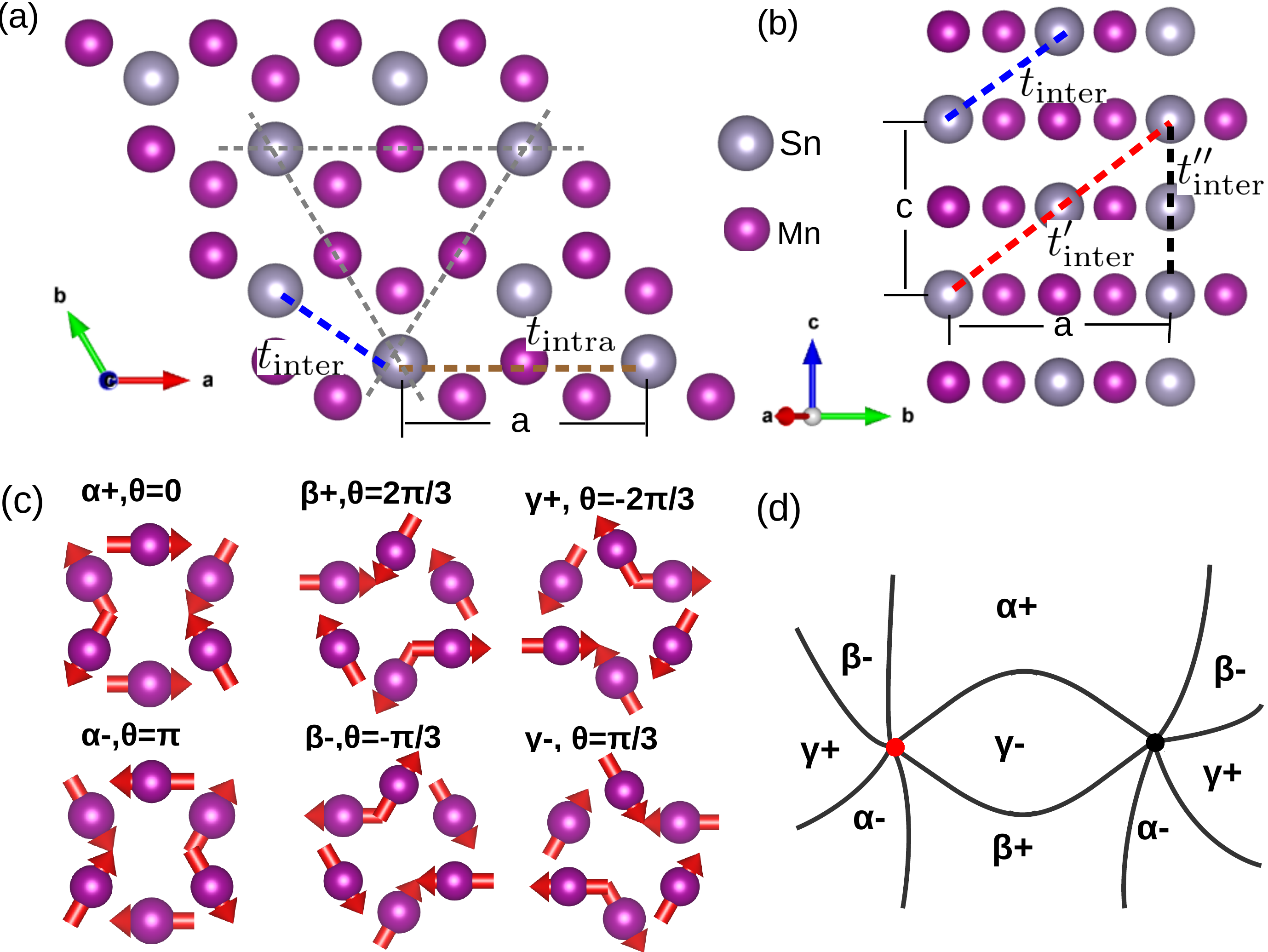}
\caption{(a) The Lattice structure of \mnsn\ from a top view, and (b) a side view. The thick dashed lines with brown, red, and 
blue colors indicate different hopping processes of the tight-binding model introduced in the text. The gray dashed lines in (a) indicate the easy axes. (c)  The six magnetic domains. (d) Schematic illustration of the $\mathbb{Z}_6$ vortex lines.
}
\label{fig:lattice}
\end{figure}

\underline{Symmetry, order parameter, and implications:}  
The \mnsn-class material crystallizes in hexagonal lattice structure with 
space group $P6_3/mmc$ as shown in Fig.~\ref{fig:lattice}(a)-(b). Taking \mnsn\ as an example,  
each Mn$^{4+}$ ion has a large spin $\sim$2-3\,$\uB$\cite{mn3sn-82-a,mn3sn-90} forming a layered Kagome lattice.

\footnote{Effects of quantum fluctuations may be neglected for such large spins, 
and hereafter we will treat these spins as classical objects.}.
The system orders antiferromagnetically in a 120$^{\circ}$ noncollinear structure as shown in Fig.~\ref{fig:lattice}(c),
with the Neel temperature $T_{N}\!\approx\!420$\,K \cite{mn3sn-75,mn3sn-82-a, mn3sn-82-b, mn3sn-90}.    
This may be understood from the hierarchy of interactions typical for
3d transition metal ions: Heisenberg exchange $J_{ij} {\bm S}_i\!\cdot
\!{\bm S}_j$ is largest, followed by Dzyaloshinskii-Moriya (DM)
interaction ${\bm D}_{ij} \cdot {\bm S}_i\!\times\!{\bm S}_j$, with
single-ion anisotropy (SIA) $-K(\hat{\bm n}_i\!\cdot\!{\bm S}_i)^2$ the weakest
effect.  The former two terms select an approximately 120$^\circ$
pattern of spins with negative vector chirality which leaves a U(1) degeneracy: any rotation of spins
within the a-b plane leaves the energy unchanged, when the SIA is
neglected.   Consequently, we can associate with these states an XY
order parameter $\psi\!=\!m_s e^{-i\theta}$, where $m_s$ is the magnitude
of the local spin moment, and $\theta$ is (minus) the angle of some specific
spin in the plane.  We focus on the ordered phase, in which $m_s$ is
uniform, and the free energy may be written in terms of $\theta$
alone.   Symmetry
dictates the form
\begin{align}
F_{\textrm{s}}=\int
  d^3\mathbf{r}\,&\Big(\,\frac{\rho}{2}\vert\bm\nabla\theta(\mathbf{r})\vert^2+\rho_1\vert\hat{\mathbf{K}}(\theta)\!\cdot\!\bm\nabla\theta\vert^2\;\nn 
&-\lambda\cos{6\theta(\mathbf{r})}
-\gamma \mathbf{B}\cdot \hat{\mathbf{K}}\,\Big)\;.
\label{eq:sine-gordon}
\end{align}
Here $\rho$ and $\rho_1$ are isotropic and anisotropic stiffnesses,
$\lambda$ is a $\mathbb{Z}_6$ anisotropy. We also introduced the XY
unit vector $\hat{\mathbf{K}}=(\cos\theta,\sin\theta,0)$, which
describes coupling $\gamma$ to a uniform magnetic field $\mathbf{B}$
(which occurs due to small in-plane canting of the moments
\cite{mn3sn-82-a, mn3sn-82-b, mn3sn-90}).  Eq.~\eqref{eq:sine-gordon}
is derived from the microscopic 
spin Hamiltonian (see Eq.~(\ref{eq:1}) in Method section), which allows
us to estimate these parameters.  We estimate
$\rho\!\approx\!0.568$\,meV/$\angstrom$,
$\rho_1\!\approx\!0.011$\,meV/$\angstrom$, and
$\lambda\!\approx\!1.159\!\times\!10^{-7}$\,meV/$\angstrom^3$ at 
temperature 50\,K.

\begin{figure}
\includegraphics[width=2.6in]{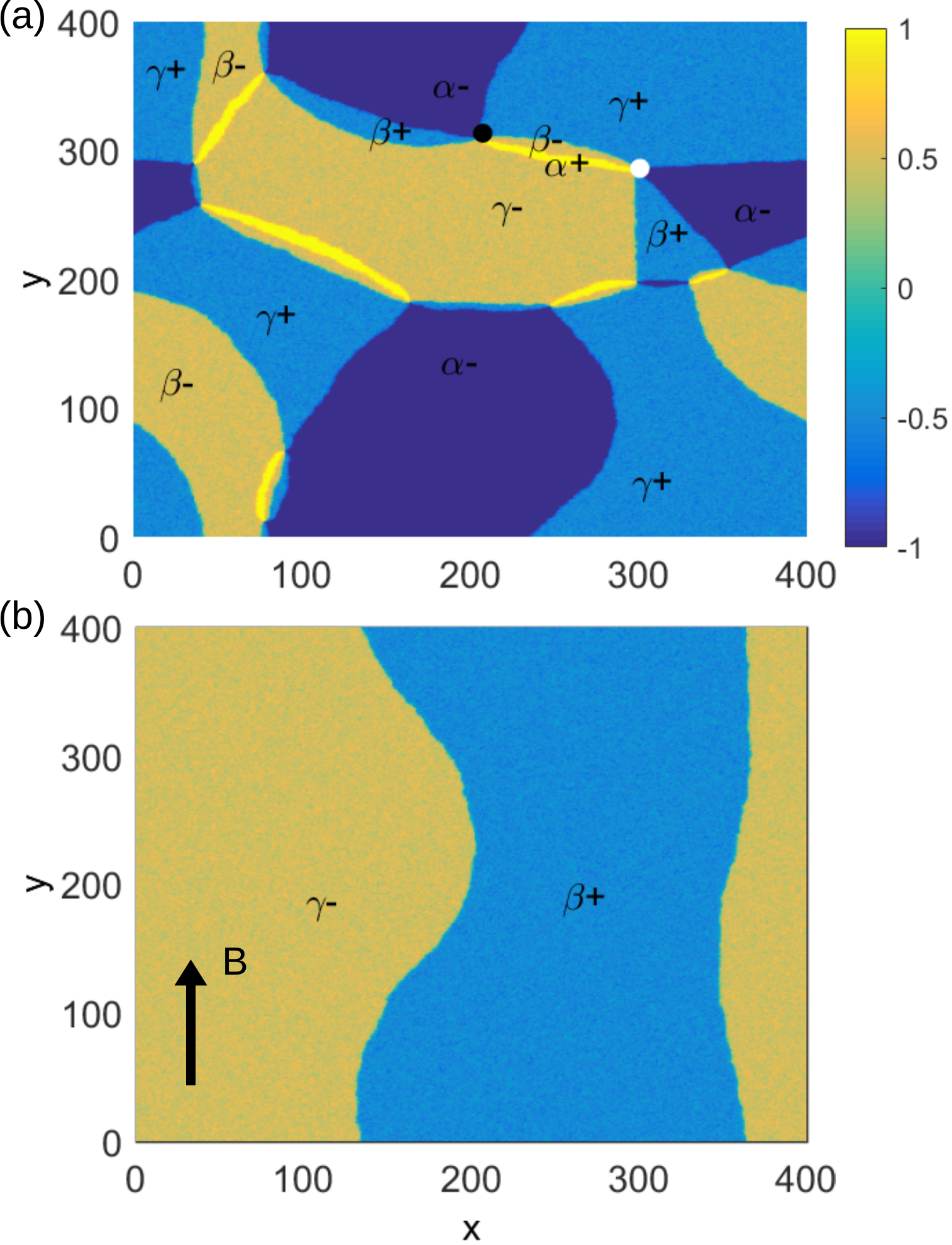}
\caption{ The spin configurations on the coarse-grained lattice at time 
$t=9600$ obtained from numerical simulations of the Langevin equation: 
(a)without any magnetic field, and (b) an external magnetic field $B=0.005\,T$ is applied along the $y$ direction.} 
\label{fig:vtdw}
\end{figure}

The $\mathbb{Z}_6$ structure of the free energy implies the existence
of six minimum energy domains in which $\theta$ maximizes
$\lambda\cos 6\theta$.  We take $\lambda>0$, for which this is
$\theta = 2\pi n/6$, with $n=0,\ldots,5$, and the corresponding spin
configurations are shown in Fig.~\ref{fig:vtdw}(c).  It is convenient to label them as
$\alpha^{+,-}$, $\beta^{+,-}$, and $\gamma^{+,-}$ as shown in
Fig.~\ref{fig:lattice}(c)., the $\pm$ superscript denoting 
domains which are time-reversal conjugates
($\theta\!\rightarrow\!  \theta+\pi$ under time-reversal).

The long-time dynamics follows from the free energy and the Langevin equation
\begin{equation}
\frac{\partial \theta(\mathbf{r},t)}{\partial t}=-\mu\frac{\delta F_{\textrm{s}}}{\delta \theta(\mathbf{r},t)}+
\mu\eta(\mathbf{r},t) + f({\bm j}),\label{eq:4}
\end{equation}
where $\eta(\mathbf{r},t)$ represents a random thermal fluctuation at
temperature $T$ obeying the Gaussian distribution of zero mean:
$\langle \eta(\mathbf{r},t)\rangle\!=\!0$, and
$\langle \eta(\mathbf{r}, t)\eta(\mathbf{r}',
t')\rangle\!=\!2k_{\textrm{B}}T\,\delta(\mathbf{r}-\mathbf{r}')\delta(t-t')$
($k_{\textrm{B}}$ is the Boltzmann constant.). $\mu$ is the damping
factor, and hereafter is set to 1. The final term $f({\bm j})$ represents non-equilibrium forces to be
discussed later.  We note that the overdamped Langevin description with a single
time derivative is valid at long times: this is sufficient for most purposes. 

Neglecting $\rho_1$ and for $\mathbf{B}=0$, Eq.~\eqref{eq:4} becomes the famous (overdamped) sine-Gordon
equation.  Its  stationary solutions include not only domains but domain walls, which are
solitons with a width $\frac{\pi}{6}\sqrt{\rho/\lambda}\!\sim\! 110$\,nm
using our estimates.  Significantly, the elementary domain walls
connect states which differ by $\Delta \theta\!=\!\pi/3$, which are {\em
  not} time-reversal conjugates.  The $\rho_1$ term leads to
orientation-dependence of the domain wall energy, and e.g. faceting of
domain boundaries.   Six of these minimal domain walls meet at curves in three
dimensions which define $\mathbb{Z}_6$ {\em vortex} lines -- see
Fig.~\ref{fig:lattice}(d), around which $\theta$ winds by $\pm 2\pi$.


To observe the microstructure predicted by the Langevin model, we
carried out a numerical simulation of a thin slab, assuming
homogeneity in the $z$ direction and discretizing the 2D continuum
model with an effective lattice constant of
$a_{\textrm{cg}}\!=\!600\angstrom$ -- see Method section for details.  
Figure ~\ref{fig:vtdw}(a) shows the spin configuration resulting from
a quench from a random initial state of a 576\,$\mu \textrm{m}^2$ sample in
zero applied field at an intermediate stage of evolution.
Clearly there are six types of domains in the figure, marked by
$\alpha^{\pm}$, $\beta^{\pm}$, and $\gamma^{\pm}$.  These sixfold
domains merge at the vortices and antivortices marked by white and
black dots respectively. 

In Fig.~\ref{fig:vtdw}(b), we show the spin configuration resulting
from the same preparation but with an applied magnetic field of
$B\!=\!0.005\,\textrm{T}$ along the [120] axis ($y$ axis).  As is
clearly shown in the figure, the field preferentially selects just two
degenerate $\beta^{+}$ ($\cos{\theta}\!=\!-1/2$) and $\gamma^{-}$
($\cos{\theta}\!=\!1/2$) domains. The orientation of the domain wall,
which tends to be normal to the [100] direction, is fixed by the anisotropic stiffness
term.    We will show that the double-domain pattern leads to a variety
of new physics including domain-wall bound states, novel transport
behavior, and domain-wall dynamics.

\underline{Minimal electronic model and electronic structure: } While the {\it ab initio} electronic
structure of \mnsn\ and Mn$_3$Ge have been studied extensively, to
study electronic properties of magnetic textures with large-scale
spatial variations and/or surface/domain wall states is impractical
with density functional theory.  Therefore we introduce a minimal 
four-band tight-binding (TB) model with a single spinor ($p_z$)
orbitals at each Sn.  As indicated by the thick dashed lines in
Fig.~\ref{fig:lattice}(a)-(b), we consider the following four hopping
processes:
\begin{align}
&t_{\textrm{intra}}(\mathbf{r}_{nm})=
t_0\,\mathbb{I}_{2\times 2}+t_{J}\,\mathbf{\sigma}\cdot\mathbf{S}_{nm}+(-1)^{\xi_{mn}}i\lambda_z\,\sigma_z\;,
\label{eq:tintra}\\
&t_{\textrm{inter}}(\mathbf{r}_{nm})=t_1\,\mathbb{I}_{2\times 2}\;,
\label{eq:tinter}\\
&t'_{\textrm{inter}}(\mathbf{r}_{nm})=
i\lambda_{\textrm{R}}\, \mathbf{e}_{\textrm{soc}}^{\mathbf{r}_{nm}}\cdot\mathbf{\sigma}\;,
\label{eq:tinterp}\\
&t''_{\textrm{inter}}(\mathbf{r}_{nm})=t_2\,\mathbb{I}_{2\times 2}\;,
\label{eq:tinterpp}
\end{align}
where the hopping from orbital $m$ centered at $\rr_m$ to orbtial $n$ centered at $\rr_n$ is
expressed as a $2\times 2$ matrix due to the spin degrees freedom of
each orbital, and $\rr_{nm}\!=\!\rr_{n}-\rr_m$.  The model includes
three spin-independent hopping terms ($t_0$ in-layer and $t_1$ and
$t_2$ inter-layer), a spin-dependent hopping $t_J$ reflecting
exchange coupling to the
Mn moment ${\bm S}$ in the middle of the bond across which the
electrons hop, and two spin-orbit coupling (SOC) terms $\lambda_z$ and
$\lambda_R$, which are important due to the heavy nature of the Sn
ions.  Details on the $\xi_{mn}$ and
$\mathbf{e}_{\textrm{soc}}^{\mathbf{r}_{nm}}$ parameters which define
the SOC are given in the Supp. Info.    Hereafter we fix the parameters of the
model as: $t_0\!=\!1$, $t_1\!=\!0.5$, $t_{J}\!=\!-0.5$,
$\lambda_z\!=\!0.5$, $t_2\!=\!-1$, and
$\lambda_{\textrm{R}}\!=\!0.2$.  We arrange ${\bm S}_{nm}$ spins to
reflect the spin order under consideration.   In the ordered state we take the spin canting angle
$\sim 1.7^{\circ}$, corresponding to a net moment $\sim 5\%$ of each
Mn spin for each Kagome cell.

\begin{figure}
\includegraphics[width=3.4in]{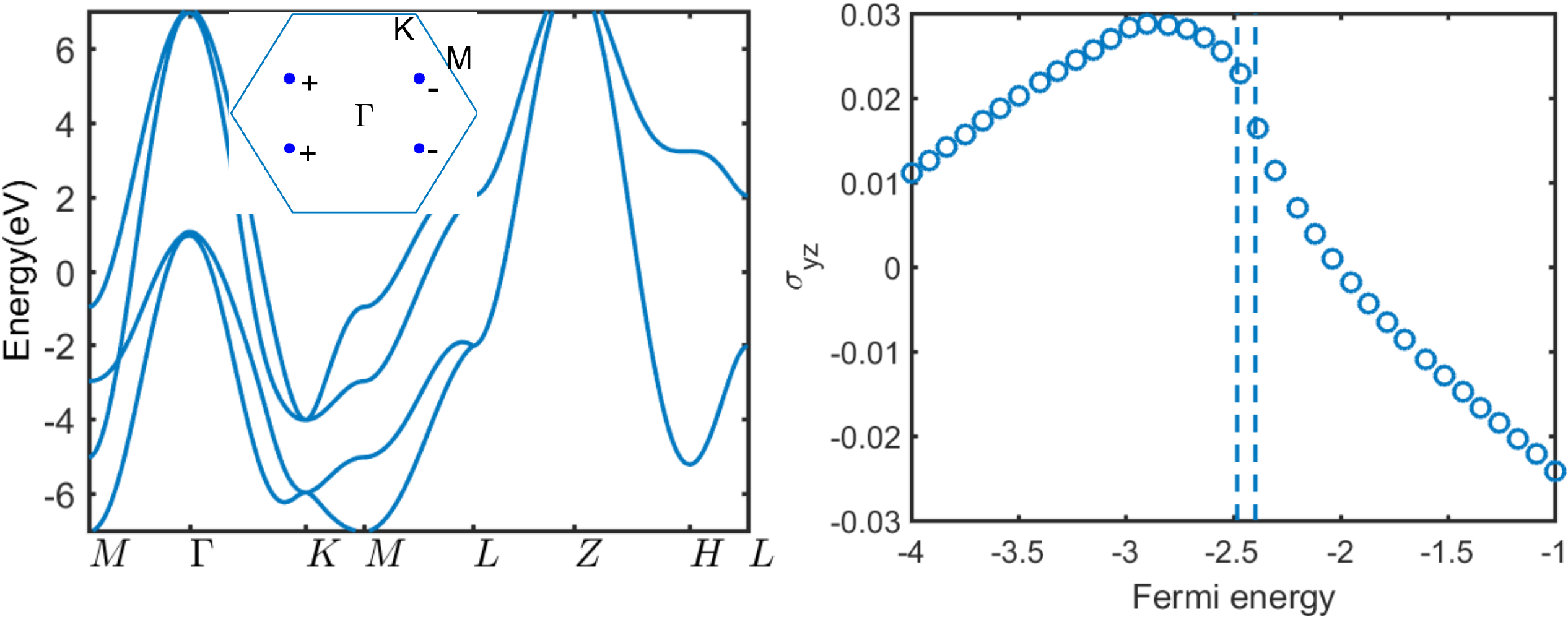}
\caption{(a) The bulk bandstructure of the tight-binding model in the $\alpha^{+}$ domain 
with $1.7^{\circ}$ spin canting. The inset indicates the positions of two different types of bulk Weyl nodes $W_1$ and $W_2$
in the $k_z\!=\!0$ plane. (b) The anomalous Hall conductivity $\sigma_{yz}$ in the $\alpha^{+}$ domain.}
\label{fig:bulk}
\end{figure}

The bulk bandstructure of the TB model  
introduced above in the $\alpha^+$ domain is shown in Fig.~\ref{fig:bulk}(a). 
We find that in the $\alpha^{+}$ domain
(see Fig.~\ref{fig:lattice}(c)), there are four Weyl
nodes at $(\pm 0.3522,0,0)$ and $(\mp 0.3522,\pm 0.3522,0)$  at energy $E_{W1}\!=\!-2.395t_0$, which
are denoted by solid blue dots in the inset of Fig.~\ref{fig:bulk}(a), with the sign corresponding to the chiralities of the Weyl nodes. There are two additional band touching points with quadratic dispersions along
the $k_z$ direction  at $(0,\pm 0.3564,0)$ at energy
$E_{W2}\!=\!-2.480t_0$. Since the dispersion is quadratic along $k_z$,
these two additional nodes carry zero Berry flux, and
do not make significant contributions to the transport properties.
The positions of the  Weyl nodes in the other five domains can be obtained by applying $C_{3z}$ and/or $\mathcal{T}$ operations to those of the $\alpha^{+}$ domain.

\underline{From magnetic structure to electronic properties:} The most
interesting feature of \mnsn\ and its relatives is the strong
influence of the magnetism on the electronic structure, and the
ability to control the latter by modifying the former.  The most basic
electronic property is the conductivity.  In the \mnsn\ family, a
symmetry analysis using crystal symmetries and Onsager relations
tightly constrains the conductivity tensor (see Sec.~\ref{sec:sigma-symm}).  In
general the antisymmetric part of the Hall conductivity is expressed
in terms of a ``Hall vector'' ${\bm Q}$, with
$\frac{1}{2}(\sigma_{\mu\nu}-\sigma_{\nu\mu}) = \frac{e^2}{2\pi
  h}\epsilon_{\mu\nu\lambda} Q_\lambda$.  We evaluate ${\bm Q}$ in a
series up to third order in the order parameter
$\psi$, and express the result in terms of $\hat{\bm K}$, which yields
\begin{equation}
  \label{eq:3}
{\bm Q}  =  q |\psi| \hat{\bm K} + \tilde{q} |\psi|^3 {\rm Im}\left[
              (\hat{K}_x + i \hat{K}_y)^3\right] \hat{\bm z}.
\end{equation}
where $q |\psi|$ and $\tilde{q}|\psi|^3$ are parameters arising from
microscopic modeling (see below).  
Since we expect the $O(|\psi|^3)$
terms to be small, we observe that the Hall vector is directed along
$\hat{\bm K}$, and lies in the xy plane.   Note that, in a Weyl semimetal with all Weyl nodes
at the Fermi level, ${\bm Q}$  is given by the fictitious dipole moment in
momentum space of the Weyl points.  While \mnsn\ is a metal and this
relation is not quantitively accurate, comparison of the Weyl nodes in the inset of
Fig.~\ref{fig:bulk}(a) shows that it is qualitatively correct.   We remark that due to the
proportionality between the Hall vector and magnetization,
${\hat{\bm K}}$ can be replaced with ${\bm M}$ in Eq.~\eqref{eq:3},
with a suitable redefinition of $q$.

To verify these symmetry considerations, we carried out a direct
calculation of the full bulk conductivity tensor of the microscopic
model using the Kubo formula (see Supplementary Information).  We show the
calculated anomalous Hall conductivity $\sigma_{zx}$ in the
  $\alpha^{+}$ domain in Fig.~\ref{fig:bulk}(b).  The result is
generically non-zero, but highly dependent upon the Fermi energy (the
horizontal axis). 

\underline{Electronic transport, electronic structure, and bound states:} The direct connection of the
conductivity to the order parameter suggests that transport can be a
fruitful probe of magnetic microstructures.  When the electronic mean
free path is shorter than the length scales of magnetic textures, a
local conductivity approximation is adequate:
${\bm J}({\bm r})\!=\!\underline{\sigma}[{\bm K}({\bm r})] {\bm
  E}({\bm r})$. From this relation and Eq.~\eqref{eq:3}, the electrostatic
potential can be determined for an arbitrary texture
$\theta(\mathbf{r})$ (see Supp. Info.), and through inversion, it should be possible to
image the magnetic domain structure purely through a
spatially-resolved electrostatic measurement.  

In the full quantum treatment, the electronic structure is
non-trivially modified by magnetic textures.  The new feature here is
the appearance of {\em Fermi arcs at domain walls}.  This is because a domain wall acts as a
sort of internal surface, at which Fermi arc states carry chiral
currents, similar to ordinary surfaces.  Without loss of generality consider a minimal energy
domain wall between the $\beta^+$ and $\gamma^-$ domains, which have
${\bm K}$ at $\pm 30^\circ$ from the $y$ axis.  The domains have Weyl
points in the $k_z\!=\!0$ plane, with chiralities that differ in the two
domains. Distinct electronic properties thus occur when this domain
wall is in an $xy$, $xz$ or $yz$ plane of the crystal.

\begin{figure}[h]
\includegraphics[width=3.4in]{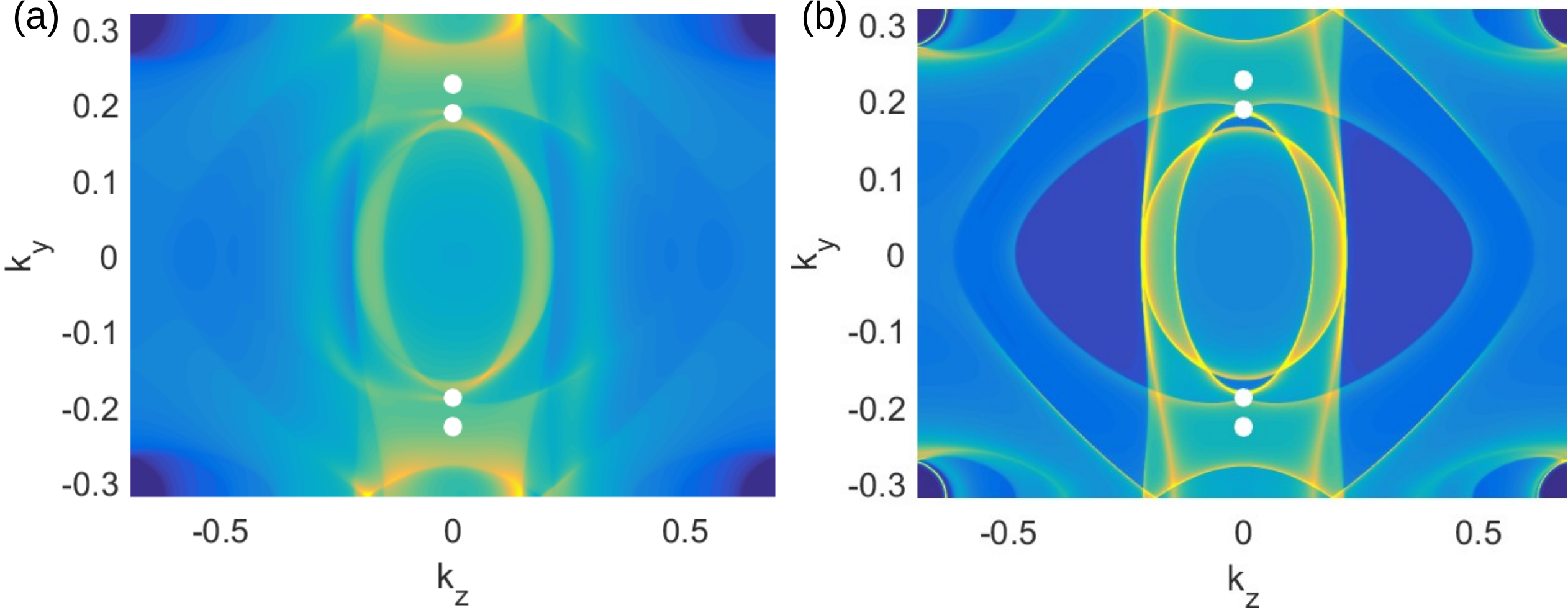}
\caption{
(a) The surface Fermi arcs of the $\beta^{+}$ domain with the surface normal vector $\hat{x}$. (b) The domain-wall Fermi arcs with the domain wall in the $yz$ plane. The white dots indicate the projection of Weyl nodes into the folded surface Brillouin zone. }
\label{fig:dwspec}
\end{figure}


Fig.~\ref{fig:dwspec}(a) shows the surface spectral functions of the
$\beta^{+}$ domain for a [100] surface.  
There are three Fermi arcs connecting the two projected Weyl nodes which are closer to the origin.  
Fig.~\ref{fig:dwspec}(c) shows the spectral function at the interface
of the  $\beta^{+}$ and $\gamma^{-}$ domains with the same
orientation.  It shows  {\it  double} the Fermi arcs found at the
interface, i.e. 6 instead of 3!  
Note that some of the projected Weyl nodes are buried in the bulk continuum due to the presence of additional Fermi surfaces 
around the Weyl nodes, which causes some of the Fermi arcs to merge
into the bulk states before connected to the Weyl nodes.
Similarly, there are also Fermi-arc states bound to the domain walls
in the $xy$ and $zx$ planes (see Supp. Info.).

Short of a challenging measurement of the momentum-resolved density of
states at a domain wall, how might one detect the presence of these
Fermi arcs and associated bound states?   We make two proposals.  First, the
in-plane transport {\em within} a domain wall may exhibit its own
anomalous Hall effect.  We checked that this indeed occurs for a 
$\beta^{+}-\gamma^{-}$ wall with $zx$-orientation, by calculating
$\sigma_{zx}$ for a supercell with two domain walls spread over 30
primitive cells.  We find $\sigma_{zx}\!=\!0.044$ for the supercell, about two times larger than
the bulk value of $0.023$ found for the same cell with a uniform
$\beta^{+}$ and $\gamma^{-}$ state and no domain walls.  This
enhancement is expected whenever $\hat{\bm K}$ is normal to the wall
in its interior.  Second, domain wall bound states can manifest as an intrinsic
resistance across the wall, since they  take away from the weight of continuum states which are
strongly transmitted and hence contribute to conductance.  We verified such a decreased conductance normal to the wall for all
domain wall orientations in numerical studies (see Supp. Info.)  

While we focused on the domain walls, it is worth noting that the
$\mathbb{Z}_6$ vortex lines may have their own electronic states.  Calculations in
the Supp. Info. show that these vortex lines show a pronounced 6-fold
pattern in their local density states, making them detectable by
scanning tunnelling microscopy \cite{stm-prb85}.

\underline{Current-driven domain-wall dynamics }
Let us now consider the feedback of the conduction electrons on the
spin texture.  This is important to control of the magnetic
microstructure electronically.    In ferromagnets, current-induced
forces on domains and domain walls have been extensively studied,
through the mechanism of spin-transfer torque\cite{stt-review-08}.
Given that the primary order parameter of the antiferromagnet is not the
magnetization, it is unclear how consideration of torque,
i.e. conservation of angular momentum, applies here.

Instead, we take a symmetry-based approach and ask how the
current ${\bm j}$ may appear as a force in the equation of motion for the  easy
spin angle $\theta$, Eq.~\eqref{eq:4}.  The result (see Supplementary
Information) is that the force takes the form
\begin{equation}
f({\bm j}) = -\sum_a \Big( p_a j_a \partial_a \theta + q_1 {\bm j}
\cdot \partial_z \hat{\bm K} + q_2 j_z {\bm\nabla}\cdot \hat{\bm K}\Big).
\label{eq:torquep}
\end{equation}
Here $p_x=p_y$, $p_z$, $q_1$ and $q_2$ are constants.  Various
arguments (see Supp. Info) suggest that $q_1$ and $q_2$, which tend to
drive the domain wall along the direction perpendicular to the current
flow, are much smaller than $p_a$, so we henceforth neglect them.

Despite the intrinsic antiferromagnetic nature of the system, the
$p_\mu$ terms appear formally very similar to spin-transfer
torques.  They could be understood in a hydrodynamic fashion as
describing ``convection'' of the spin texture with or against the
current flow: indeed added to Eq.~\eqref{eq:4} , these terms are
equivalent to a Galilean boost and consequently velocity $v_a =
\mu p_a j_a$.   This leads to concrete experimental proposals.
Specifically, in the geometry of Fig.~\ref{fig:vtdw}(b), a current
applied along the $x$ direction controls the position of the wall.
The non-dissipative Hall voltage measured between two contacts across the $y$ direction at
fixed $x$ can thereby be {\em switched} by purely electrical means, as
the domain wall moves to the left or right of the contacts.

The results of this paper provide the framework to design and model
the spin dynamics and topologically-influenced electrical transport in
the negative vector chirality
antiferromagnets \mnsn\ and Mn$_3$Ge, and the methodology may be
applied more broadly to XY-like antiferromagnetic systems.   Weyl
nodes in the electronic structure induce Fermi arc bound states that
influence transport in the presence of domain walls.  In addition to advancing the fundamental physics of Weyl fermions in
noncollinear antiferromagnets, these results mark the \mnsn-class of materials as
promising candidates for novel magnetic storage devices.

\section{Methods}

\subsection{Derivation of the sine-Gordon model}
\label{sec:sg1}

In this section, we present a derivation of the continuum sine-Gordon
energy from a microscopic spin Hamiltonian.  We consider the following
spin interactions
\begin{align}
  \label{eq:1}
  H_{\textrm{s}}  = & J_1 \sum_{\langle ij\rangle_{xy}} \mathbf{S}_i
                       \cdot \mathbf{S}_j + J_2 \sum_{\langle
                       ij\rangle_z} \mathbf{S}_i
                       \cdot \mathbf{S}_j  \nn
                       &  + \sum_{\langle ij\rangle_{xy}} \mathbf{D}_{ij}
                           \cdot \mathbf{S}_i\times\mathbf{S}_j -
                           \sum_i K \left( \hat{\mathbf{n}}_i \cdot \mathbf{S}_i\right)^2.
\end{align}
Here we indicated a sum over nearest-neighbors in the $xy$ plane by
$\langle ij\rangle_{xy}$ and similarly nearest-neighbors in successive
$xy$ planes by $\langle ij\rangle_{z}$.  The spin $\mathbf{S}_i$ is considered
as a classical vector with fixed length $m_s$.
The positive constants
$J_1,J_2$ are isotropic Heisenberg interactions on these bonds.  We
include a Dzyaloshinskii-Moriya interaction specified by the
D-vector $\mathbf{D}_{ij}=-\mathbf{D}_{ji}$ which takes the form allowed 
by the symmetry of the kagom\'e lattice.  
Specifically, if we choose $i$ and $j$ for a given bond so
that $i\rightarrow j$ proceeds {\em counter-clockwise} on the triangle to
which the bond belongs, then we have
\begin{equation}
  \label{eq:2}
  \mathbf{D}_{ij} = D \hat{\mathbf{z}} + D' \hat{\mathbf{z}} \times \hat{\mathbf{e}}_{ij},
\end{equation}
where $\hat{\mathbf{e}}_{ij}$ is the unit vector oriented from site
$i$ to site $j$.  It worth to note that prior modeling of the spin
interactions in Mn$_3$Sn have included the $D$ term but not the $D'$ one.  The $K$ term is a local easy-axis anisotropy, which
is determined by the unit vector $\hat{\mathbf{n}}_i$  oriented along
the direction between the spin $i$ and either of its nearest Sb ions
as indicated by the gray dashed line in Fig.~\ref{fig:lattice}(a).

We first assume a uniform spin configuration, which is sufficient to describe
the ground state, and it determines the $Z_6$ anisotropy $\lambda$.  
There are six spins per unit
cell, which form two triangles, one in each of the two distinct
layers.  By inspection, we find that there is inversion symmetry 
in the magnetic ground state and we only have to consider the spins in one
triangle. The Heisenberg term in Eq.~(\ref{eq:1}) is minimized by requiring the three spins
to lie in a plane at 120 degree angles to one another.  
The plane of the spins is undetermined by the Heisenberg
term, but fixed by the DM interaction.  
To leading order in the DM 
terms, the ground state is of the form
\begin{equation}
  \label{eq:10}
  \mathbf{S}^{(0)}_a = m_s\begin{pmatrix}
\cos (-\theta - \tfrac{2\pi a}{3})\\
\sin(-\theta - \tfrac{2\pi a}{3})\\
0
\end{pmatrix}.
\end{equation}
We also include small ``canting'' of the spins away from the rigid
configuration.  Formally, we do this by writing $K\rightarrow \eta K$,
$D\rightarrow \eta D$, $D'\rightarrow \eta D'$, and carrying out
perturbation theory in $\eta$.  To do so, we let
\begin{equation}
  \label{eq:11}
  \mathbf{S}_a = m_s\begin{pmatrix}
\sqrt{1-u_a^2}\cos (-\theta - \tfrac{2\pi a}{3}+\phi_a)\\
\sqrt{1-u_a^2}\sin(-\theta - \tfrac{2\pi a}{3}+\phi_a )\\
u_a
\end{pmatrix},
\end{equation}
where $a=1,2,3$ denote the sublattice indices of the kagome\' lattice.
We set $\phi_3 = -\phi_1-\phi_2$ to keep $\phi_1,\phi_2$ linearly
independent of $\theta$.  We also write $u_{1,2,3}$ and $\phi_{1,2}$
in a series in $\eta$,
$u_a = \sum_{n=1}^\infty u_{a,n} \eta^n$, 
  $\phi_a =  \sum_{n=1}^\infty \phi_{a,n}\eta^n.$
Inserting Eq.~\eqref{eq:11} into the spin Hamiltonian
, we then obtain a formal expansion of the energy order by
order in $\eta$.
Keeping the expansion to the third order in $\eta$, then minimizing with respect to $u_{a}$
and $\phi_a$,  we obtain the optimal spin configuration to first order in the canting angles,
and the ground state energy to third order in $\eta$:
\begin{equation}
  \label{eq:13}
  E_{\textrm{gs}} = \mathcal{E}_0 - \frac{K^3}{12(J_1+J_2)^2} m_s^2 \cos 6\theta,
\end{equation}
where $\mathcal{E}_0$ is a $\theta$-independent constant.
The coefficient of $\cos 6\theta$
in Eq.~\eqref{eq:13} allows us to determine $\lambda$ in the
sine-Gordon model.  

Further results are obtained by adding the effect of a Zeeman magnetic
field to the energy.
We repeat the previous analysis, taking the magnetic field $\mathbf{B} \rightarrow \eta
\mathbf{B}$ as well.  This corresponds to considering the Zeeman energy much
small than $J_1+J_2$, an excellent approximation.  It turns out that  the leading term in the
in-plane magnetization  is
\begin{equation}
  \label{eq:17}
  \mathbf{M}_{xy} = \frac{K g m_s}{J_1 + J_2} \begin{pmatrix}
    \cos\theta \\
 \sin \theta \\
0 \end{pmatrix} \equiv \frac{K g m_s}{J_1 + J_2}  \hat{\mathbf{K}},
\end{equation}
One may note that the angle of the net magnetization $\theta$ is {\em minus}
the $U(1)$ rotation angle , which is due to the antichiral spin texture on
kagome\' lattice. We refer the readers to Supp. Info. for more details.

The out-of-plane
magnetization turns out to be parametrically smaller by a factor of $D'/J$:
\begin{equation}
  \label{eq:18}
  M_{z} = -\frac{D' K g m_s}{\sqrt{3}(J_1+J_2)^2} \sin 3\theta.
\end{equation}
The above equation shows that the magnetization does not stay entirely
within the $xy$ plane.  For $\lambda>0$, where the minimum energy
values of $\theta$ are multiples of $2\pi/6$, then $\sin 3\theta = 0$
and the bulk $z$-axis magnetization within a uniform domain vanishes.
This corresponds to the case in which one of the three spins on each
triangle orients along its easy axis, directly toward a neighboring
Sn.  One can verify that this situation preserves a mirror plane which
enforces $M^z=0$.  For $\lambda<0$, however, $\sin 3\theta = \pm 1$ at
the minimum values of $\theta$, and so the domains are expected to
have a small bulk magnetization, reduced by a factor of
$D'/\sqrt{3}(J_1+J_2)$ relative to the in-plane magnetization.  Since
such a $z$-axis magnetization seems not to have been detected in
\mnsn, we take this as evidence in favor of the $\lambda>0$ state.
Even for this state, however, we see that the out of plane
magnetization $M^z$ becomes non-zero within domain walls.  We remark
in passing that experiments show that in Mn$_3$Ge the anomalous Hall
conductivity within the $xy$ plane is small but
nonvanishing\cite{mn3ge-exp}, suggesting that the $\lambda>0$
state is realized in Mn$_3$Ge.

We continue to study the magnetic susceptibilities in the high-field regime, i.e., when the spontaneous magnetization is much smaller than the field-induced one. When the field is within the $xy$ plane, 
$\mathbf{B} = B(\cos{\alpha},\sin{\alpha},0)$, the in-plane susceptibility
is expressed as 
\begin{equation}
\chi_{\perp c}\!=\!\chi_{\perp c, 0}+\chi_{\perp c, 1}\cos{6\alpha}, 
\end{equation}
where
\begin{eqnarray}
\chi_{\perp c, 0} & = & \frac{g^2}{J_1+J_2}\Big(\,1 - \frac{\sqrt{3}D}{J_1+J_2}\,\Big)\;,
\label{eq:chi0}\\
  \chi_{\perp c,1} & = & \frac{Kg^2}{6(J_1+J_2)^2}.
\label{eq:chi1}
\end{eqnarray}
It follows that the in-plane magnetization is linear in field with an offset $M_{xy}$ (see Eq.~(\ref{eq:17})), and a six-fold modulation.  Measurement of the six-fold modulation
provides a way to determine $K/(J_1+J_2)$. 

On the other hand, when the magnetic field is along the $z$ direction, the out-of-plane
susceptibility is expressed as
\begin{equation}
\chi_{\parallel c}=\frac{g^2}{J_1+J_2}\Big(\,1 - \frac{\sqrt{3}D}{3(J_1+J_2)}\,\Big)\;.
\label{eq:chiz}
\end{equation}
The exchange $J_1+J_2$ and DM parameter $D$ can be determined by susceptibility measurements
using Eq.~(\ref{eq:chi1}) and (\ref{eq:chiz}).

To obtain the full continuum theory, we
need to allow slow spatial variations of $\theta$.  To do so, we
introduce the parametrization similar to Eq.~\eqref{eq:11} but with no assumptions
about uniformity or symmetry:
\begin{equation}
  \label{eq:24}
    \mathbf{S}_{a,s}(\mathbf{r}) = m_s\begin{pmatrix}
\sqrt{1-u_{a,s}^2(\mathbf{r})}\cos (-\theta(\mathbf{r}) - \tfrac{2\pi a}{3}+\phi_{a,s}(\mathbf{r}))\\
\sqrt{1-u_{a,s}^2(\mathbf{r})}\sin(-\theta(\mathbf{r}) - \tfrac{2\pi a}{3}+\phi_{a,s}(\mathbf{r}) )\\
u_{a,s}(\mathbf{r}).
\end{pmatrix}
\end{equation}

The idea now is to insert the ansatz in Eq.~\eqref{eq:24} into
the spin Hamiltonian, and expand both in powers of $\phi_{a,s}$ and
$u_{a,s}$ {\em and} in gradients.    
The leading stiffness terms can be obtained by minimizing the spin Hamiltonian with respect to $\phi_{a,s}$ and $u_{a,s}$ at fixed $\theta$.  The result is
\begin{eqnarray}
  \label{eq:27}
  H & = \sum_{\mathbf{r}} & \Big[ \frac{9 m_s^2 a_0^2 J_1
    (3J_1+2J_2)}{2(3J_1+J_2)} \left( |\partial_x \theta|^2 +
          |\partial_y\theta|^2\right) \nonumber \\
  && + 12 d^2 m_s^2 J_2 |\partial_z \theta|^2 \Big].
\end{eqnarray}
From this the stiffnesses can be read off.

The only remaining term in the continuum energy to be discussed is
anisotropic gradient one.  For simplicity, we neglect the possible
effect of the DM interactions on this term, and set $D=D'=0$.   We  anticipate that the
anisotropic stiffness appears with a coefficient of order $K$.  We
treat the gradients, small canting angle, and $K$, all of the same
order, and expand the energy up to $O(\phi^3)$.  Due to
the lack of mixing between $z$ and $xy$ components when $D'=0$, the out of
plane canting components $u_{a,s}$ have no effect and we can set them
to zero.  Then one may minimize the energy with respect to $\phi_a$, 
and select the terms second
order in gradients.  After carrying through this algebra, we obtain
  \begin{equation}
    \label{eq:28}
    H^{(3)} =   \rho_1 \sum_{\mathbf{r}} \Big[
    (\hat{\mathbf{K}}\cdot \mathbf{\nabla}\theta)^2 -\tfrac{1}{2} |\mathbf{\nabla}_\perp \theta|^2\Big],
  \end{equation}
with $\hat{\mathbf{K}} = (\cos\theta,\sin\theta,0)$ and
\begin{equation}
  \label{eq:29}
  \rho_1 = \frac{15 m_s^2 a_0^2 K J_1 (9J_1^2 + 3J_1 J_2 +
      2J_2^2)}{2(J_1+J_2)(3J_1+J_2)^2} .
\end{equation}
We refer the readers to Supp. Info. for more details about the derivation of the continuum theory.

\subsection{Evaluations of the sine-Gordon parameters}
\label{sec:sg2}

As discussed above, the exchange interaction $J_1+J_2$ and DM interaction $D$ can be determined by the susceptibility measurements using Eq.~(\ref{eq:chi0}) and (\ref{eq:chiz}).
We have used the data measured at $300$\,K as reported in Ref.~\onlinecite{mn3sn-exp},
and find that $J_1+J_2=5.606$\,meV, and $D=0.635\,$meV. 
The ratio $K/(J_1+J_2)$ can be determined by
measuring the six-fold modulations of the in-plane magnetizations
\footnote{We thank the groups of Professors Satoru Nakatsuji and Yoshichika Otani for sharing their unpublished data.}.
It turns out that $K=0.187$\,meV. The finite-temperature effect is taken into account
by letting $m_s\to m_s(T)$, where $m_s(T)$ is the mean-field expectation value of a spin $1$ at temperature $T$. In particular, at $T=50$\,K, $m_s(T)=0.92$. Given the specific
values of $J_1+J_2$, $D$, $K$ and $m_s(T)$, we evaluate the $\mathbb{Z}_6$ anisotropy
$\lambda=1.159\times 10^{-7}$\,meV/$\angstrom^{-3}$, the isotropic stiffness $\rho=0.568$\,meV/\angstrom, and the anisotropic stiffness $\rho_1=0.011$\,meV/\angstrom. We may also
obtain the canting moment from Eq.~(\ref{eq:17}), which turns out to be 0.061\,$\mu_{\textrm{B}}$ per unit cell at 50\,K, from which we obtain the Zeeman energy density for in-plane magnetic field $B$ as $h=M_{xy}B=2.814\times 10^{-3}B$\,meV$\angstrom^{-3}$T$^{-1}$, where $B$ is the magnitude of the magnetic field in units of Tesla. Note that the estimated canting moment
is about 5 times larger than the experimental measurements, and we suspect that the 
measured value has underestimated the canting moment due to the cancellation from different domains.


\subsection{Domain-wall bound states}
\label{sec:dw}

The surface spectral functions as shown in Fig.~\ref{fig:dwspec}(a)
are calculated using the method proposed in Ref.~\onlinecite{surface-gf}.
In order to calculate the domain-wall spectral functions, we include the domain-wall
layers coupled to the two semi-infinite domains, and the thickness of the domain wall is $N_{\textrm{dw}}$ (in units of lattice constants). The spins vary smoothly from one domain to the other across the domain wall.  The domain-wall spectral function can be solved using the Dyson equation,
\begin{equation}
G_{\textrm{dw}}=G_{\textrm{dw}}^{0}+
G_{\textrm{dw}}^{0}(\mathbf{k},\omega)\Sigma_{\textrm{dw}} G_{\textrm{dw}}\;,
\end{equation}
where $G_{\textrm{dw}}$ represents the retarded Green's function of the domain wall
including the effects due to the couplings to two domains, while $G_{\textrm{dw}}^{0}$ is 
the ``bare" Green's function excluding the coupling between the domain wall and the domains,
and $\Sigma_{\textrm{dw}}$ is the self energy from the coupling. In the above equation the dependence on
the 2D wavevector $\mathbf{k}$ and the frequency $\omega$ is implicit.
More specifically,
\begin{equation}
G_{\textrm{dw}}^{0}=\begin{pmatrix}
G_{\beta^+}^s & 0 &  0\\  
0 &  G_{00} & 0 \\
0  &  0 &  G_{\gamma^-}^s
\end{pmatrix}\;,
\end{equation}
where $G_{00}$ is the Green's function of the isolated domain-wall layers, and $G_{\beta^+}^s$ and
$G_{\gamma^-}^s$ denote the surface Green's functions of the $\beta^+$ and $\gamma^-$ domains
calculated using the iterative scheme proposed in Ref.~\onlinecite{surface-gf}. The self energy
$\Sigma_{\textrm{dw}}$ is simply the coupling between the domain wall and the domains,
\begin{equation}
\Sigma_{\textrm{dw}}=\begin{pmatrix}
0 & H_{\beta^+,\textrm{dw}} & 0\\
H_{\beta^+,\textrm{dw}}^{\dagger} & 0 & H_{\textrm{dw},\gamma^-} \\
0 & H^{\dagger}_{\textrm{dw},\gamma^-} &  0\;.
\end{pmatrix}
\end{equation}

\subsection{Symmetry analysis on the conductivity tensor}
\label{sec:sigma-symm}
In this section we derive the symmetry-allowed expressions of the bulk conductivity tensor. 
We consider four generators of the symmetry operations of space group $P6_3/mmc$: $120^{\circ}$ rotation about $z$
axis $C_{3z}$, $180^{\circ}$ rotation about an in-plane axis which is parallel to [100] and half-way between the $z\!=\!0$
and $z\!=\!c/2$ plane $C_{2x}$, a $180^{\circ}$ screw rotation about $z$ axis $C_{2z}^{s}$, and finally inversion $\mathcal{P}$.
The full conductivity tensor $\underline{\sigma}$ can be expressed as:
\begin{align}
\underline{\sigma}=&\underline{\sigma}^0+\sum_{\mu=x,y}\underline{A}^{\mu}\vert\psi\vert\hat{K}_{\mu} +\sum_{\mu,\nu}\underline{B}^{\mu\nu}\vert\psi\vert^2\hat{K}_{\mu}\hat{K}_{\nu}+\mathcal{O}(\vert\psi\vert^3)\;,
\label{eq:sigma}
\end{align}
where $K_{\mu}$ is the $\mu$th component of the order parameter $\mathbf{\hat{K}}=(\cos\theta,\sin\theta,0)$.
$\underline{\sigma}^{0}$ is the term which is independent of magnetic state,  while $\underline{A}^{\mu}$ and $\underline{B}^{\mu\nu}$ couples to  $\mathbf{\hat{K}}$ to the linear and quadratic orders
respectively. Due to Onsager reciprocal relation, the terms which
are odd (even) in $\hat{K}_{\mu}$ have to be antisymmetric (symmetric). Thus $\underline{A}^{\mu}$ is antisymmetric,
while $\underline{\sigma}^{0}$ and $\underline{B}^{\mu\nu}$ are symmetric. 
\begin{table}
\setlength{\tabcolsep}{4pt}
\caption{Symmetry representations}
\begin{ruledtabular}
\begin{tabular}{lclclclc}
$g$  & $\Gamma_g$ & $\mathcal{O}_g$\\
\hline
$C_{3z}$  &  $\begin{pmatrix}  -\frac{1}{2} & -\frac{\sqrt{3}}{2} \\ \frac{\sqrt{3}}{2} &  -\frac{1}{2} \end{pmatrix}$ 
& $\begin{pmatrix}  -\frac{1}{2} & -\frac{\sqrt{3}}{2} & 0 \\ \frac{\sqrt{3}}{2} &  -\frac{1}{2} &  0 \\ 0&0&1 \end{pmatrix}$   \\
\vspace{6pt}
$C_{2x}$  &  $\begin{pmatrix} 1 & 0 \\ 0 &  -1 \end{pmatrix}$   &  $\begin{pmatrix}  1 & 0 & 0 \\ 0 &  -1 &  0 \\ 0&0&-1
\end{pmatrix}$   \\
\vspace{6pt}
$C_{2z}^{s}$ & $\begin{pmatrix} -1 & 0 \\ 0 &  -1 \end{pmatrix}$ & $\begin{pmatrix}  -1 & 0 & 0 \\ 0 &  -1 &  0 \\ 0&0&1
\end{pmatrix}$ \\
\vspace{6pt}
$\mathcal{P}$ &  $\begin{pmatrix} 1 & 0 \\ 0 &  1 \end{pmatrix}$&$\begin{pmatrix}-1 & 0 & 0 \\ 0 &-1 &0 \\0&0&-1\end{pmatrix}$  
\end{tabular}
\end{ruledtabular}
\label{table:symm}
\end{table}
The conductivity tensor should be invariant under a symmetry operation $g$, which means
\begin{align}
&\mathcal{O}_{g}\,\underline{\sigma}^{0}\,\mathcal{O}_{g}^{T}=\underline{\sigma}^{0}\;\nn
&\mathcal{O}_{g}\,\underline{A}^{\mu}\,\mathcal{O}_{g}^{T}=\sum_{\mu=x,y}\Gamma_{g,\mu\nu}\,\underline{A}^{\nu}\;\nn
&\mathcal{O}_{g}\,\underline{B}^{\mu\nu}\,\mathcal{O}_{g}^{T}=
\sum_{\mu',\nu'}\Gamma_{g,\mu\mu'}\Gamma_{g,\nu\nu'}\,\underline{B}^{\mu'\nu'}\;,
\label{eq:sigma_sym}
\end{align}
where $\mathcal{O}_g$ is a $3 \times 3$ matrix representing the symmetry operation $g$ on a 3D real vector, 
and $\Gamma_{g}$ is a $2\times 2$ matrix  representing the symmetry operation $g$ acting on the $xy$
component of $\mathbf{\hat{K}}$.
The symmetry representations $\mathcal{O}_g$ and $\Gamma_g$ are tabulated in Table~\ref{table:symm}.
After solving 
Eq.~(\ref{eq:sigma_sym}), we obtain the symmetry-allowed conductivity tensor:
\begin{align}
\sigma_{\mu\nu}= \sigma_\parallel \delta_{\mu\nu} +(\sigma_\perp - \sigma_\parallel)
\delta_{\mu z}\delta_{\nu z} + q \vert\psi\vert\epsilon_{\mu\nu\lambda} \hat{K}_\lambda + b_1\vert\psi\vert^2\hat{K}_\mu \hat{K}_\nu\;,
\label{sigma-full}
\end{align}
where $\sigma_{\perp}$ denotes the out-of-plane diagonal conductivity, $\sigma_{\parallel}$ denotes
the isotropic part of the in-plane diagonal conductivity, $b_1$ denotes the anisotropic part of
the in-plane conductivity, and finally $q$ term denotes the  anomalous Hall conductivity.
We refer the readers to Supp. Info. for more details about the numerical calculations of the conductivities.

\subsection{Symmetry analysis on the spin-transfer torques}
\label{sec:f-symm}
In this section we provide a derivation of Eq.~(\ref{eq:torquep}).
The general expression for the current-induced  spin-transfer torque is:
\begin{align}
f(\mathbf{j})=-\sum_{a,b}(B_0^{ab}\,j_{a}\,\partial_b\theta + B_{1}^{ab}\,j_{a}\,\partial_b\theta\cos{\theta}+B_{2}^{ab}\,j_{a}\,\partial_b\theta\sin{\theta})\;,
\label{eq:f}
\end{align}
where $j_a$ denotes the $a$ component of the electric current with $a,b\!=\!x,y,z$. 
The dependence of $\theta$ on position and time is implicit. 
It is convenient to decompose Eq.~(\ref{eq:f}) as $f(\mathbf{j})=f_p(\mathbf{j})+f_{q}(\mathbf{j})$,
where the leading term $f_p(\mathbf{j})=-\sum_{a,b} B_0^{ab}j_{a}\partial_b\theta$, and the subleading term $f_{q}(\mathbf{j})=-\sum_{a,b}(B_{1}^{ab}\,j_{a}\,\partial_b\theta\cos{\theta}+B_{2}^{ab}\,j_{a}\,\partial_b\theta\sin{\theta})$.

Let us first consider the $f_p(\mathbf{j})$ term. The $U(1)$ rotation $\theta$ is transformed to $\theta_g$
after a symmetry operation $g$. More specifically,
\begin{align}
&C_{3z}: \hspace{6pt} \theta\to\theta_g=\theta+\frac{2\pi}{3}\;,\nn
&C_{2x}:\hspace{6pt} \theta\to\theta_g=-\theta\;,\nn
&C_{2z}^{s}: \hspace{6pt} \theta\to\theta_g=\theta+\pi\;,\nn
&\mathcal{P}: \hspace{6pt} \theta\to\theta_g=\theta\;.
\end{align}
With a symmetry operation $g$, the spin-transfer torque $f_p(\mathbf{j})\!\to\!f_{p,g}(\mathbf{j})$,
where
\begin{align}
f_{p,g}(\mathbf{j})&=-\sum_{a,a',b,b'}\,B_0^{ab}\,\mathcal{O}_{g,a'a}^{T}\,\mathcal{O}_{g,bb'}
\,j_{a'}\,\partial_{b'}\,\theta_g\;\nn
&=-\sum_{a,b}B_0^{ab}\,j_a\,\partial_b\,\theta_g\;.
\end{align}
It follows that
\begin{equation}
\sum_{a'b'}\,\mathcal{O}_{g,aa'}^{T}\,B_{0}^{a'b'}\,\mathcal{O}_{g,b'b}=B_0^{ab}\;,
\label{eq:eom-symm1}
\end{equation}
where $\mathcal{O}_g$ is tabulated in Table~\ref{table:symm}.
After solving Eq.~(\ref{eq:eom-symm1}), one obtains the $p_a j_a\partial\theta$ term in Eq.~\eqref{eq:torquep}. 

The $f_q(\mathbf{j})$ term is more complicated. 
Under a symmetry transformation $g$,  $f_q(\mathbf{j})\to f_{q,g}(\mathbf{j})$, where
\begin{widetext}
\begin{align}
f_{q,g}(\mathbf{j})&=
-\sum_{aa'bb'}\sum_{\mu=x,y}\Big(\,B_1^{ab}\,\mathcal{O}^{T}_{g,a'a}\,\mathcal{O}_{g,bb'}\,j_{a'}\,\partial_{b'}\theta_g\,\Gamma_{g,x\mu}\,\hat{K}_{\mu} +
B_2^{ab}\,\mathcal{O}^{T}_{g,a'a}\,\mathcal{O}_{g,bb'}\,j_{a'}\,\partial_{b'}\theta_g\,\Gamma_{g,y\mu}\,\hat{K}_{\mu}\,\Big)\;\nn
&=-\sum_{ab}\Big(\,B_1^{ab}\,j_{a}\,\partial_{b}\theta_g\,\hat{K}_x+B_2^{ab}\,j_{a}\,\partial_{b}\theta_g\,\hat{K}_y\,\Big)\;,
\end{align}
\end{widetext}
where the matrix $\Gamma_g$ has been tabulated in Table~\ref{table:symm}, $\hat{K}_x=\cos{\theta}$, and $\hat{K}_y=\sin{\theta}$.  
From the above equation it follows
that
\begin{align}
\sum_{a'b'}\mathcal{O}_{g,aa'}^{T}\,(\,B_1^{a'b'}\,\Gamma_{g,xx}+B_2^{a'b'}\,\Gamma_{g,yx}\,)\,\mathcal{O}_{g,b'b}
=B_1^{ab}\;,\nn
\sum_{a'b'}\mathcal{O}_{g,aa'}^{T}\,(\,B_2^{a'b'}\,\Gamma_{g,yy}+\,B_1^{a'b'}\,\Gamma_{g,xy}\,)\,\mathcal{O}_{g,b'b}
=B_2^{ab}.\;
\end{align}
One would obtain the $q_1$, $q_2$ terms in Eq.~(\ref{eq:torquep}) after solving the above equations.

\bibliography{mn3sn_short}

\underline{Acknowledgements:}  We thank the groups of Professors
Satoru Nakatsuji and Yoshichika Otani for introducing us to these
materials and sharing their data.  This research was supported by the
National Science Foundation under grant number DMR1506119.
 
\end{document}